\documentclass[prl,superscriptaddress,showpacs,amssymb,amsmath,amsfonts,aps]{revtex4}

\usepackage{graphicx}

\def\be{\begin{equation}}
\def\ee{\end{equation}}
\def\bea{\begin{eqnarray}}
\def\eea{\end{eqnarray}}

\begin{document}

\title{The Onset of Quark-Hadron Duality in Pion Electroproduction}
\author{
T.~Navasardyan,$^{1}$ 
G.S.~Adams,$^{2}$ A.~Ahmidouch,$^{3}$ T.~Angelescu,$^{4}$
J.~Arrington,$^{5}$ R.~Asaturyan,$^{1}$ O.K.~Baker,$^{6,7}$
N.~Benmouna,$^{9}$ C.~Bertoncini,$^{10}$
H.P.~Blok,$^{11}$ W.U.~Boeglin,$^{12}$ P.E.~Bosted,$^{13}$ H.~Breuer,$^{8}$
M.E.~Christy,$^{6}$ S.H.~Connell,$^{14}$ Y.~Cui,$^{15}$
M.M.~Dalton,$^{14}$ S.~Danagoulian,$^{3}$ D.~Day,$^{16}$
T.~Dodario,$^{15}$ J.A.~Dunne,$^{17}$ D.~Dutta,$^{18}$
N.~El~Khayari,$^{15}$ R.~Ent,$^{7}$ H.C.~Fenker,$^{7}$
V.V.~Frolov,$^{19}$ L.~Gan,$^{20}$ D.~Gaskell,$^{7}$
K.~Hafidi,$^{5}$ W.~Hinton,$^{6,7}$ R.J.~Holt,$^{5}$
T.~Horn,$^{8}$ G.M.~Huber,$^{21}$ E.~Hungerford,$^{15}$
X.~Jiang,$^{22}$ M.~Jones,$^{7}$ K.~Joo,$^{23}$
N.~Kalantarians,$^{14}$ J.J.~Kelly,$^{8}$ C.E.~Keppel,$^{6,7}$
V.~Kubarovski,$^{2}$ Y.~Li,$^{15}$
Y.~Liang,$^{24}$ S.~Malace,$^{4}$
P.~Markowitz,$^{12}$ E.~McGrath,$^{25}$ P.~McKee,$^{16}$
D.G.~Meekins,$^{7}$ H.~Mkrtchyan,$^{1}$ B.~Moziak,$^{2}$
G.~Niculescu,$^{16}$ I.~Niculescu,$^{25}$ A.K.~Opper,$^{24}$
T.~Ostapenko,$^{26}$ P.~Reimer,$^{5}$ J.~Reinhold,$^{12}$
J.~Roche,$^{7}$ S.E.~Rock,$^{13}$ E.~Schulte,$^{5}$
E.~Segbefia,$^{6}$ C.~Smith,$^{16}$ G.R.~Smith,$^{7}$
P.~Stoler,$^{2}$ V.~Tadevosyan,$^{1}$ L.~Tang,$^{6,7}$
M.~Ungaro,$^{2}$ A.~Uzzle,$^{6}$ S.~Vidakovic,$^{21}$
A.~Villano,$^{2}$ W.F.~Vulcan,$^{7}$ M.~Wang,$^{13}$
G.~Warren,$^{7}$ F.~Wesselmann,$^{16}$ B.~Wojtsekhowski,$^{7}$
S.A.~Wood,$^{7}$ C.~Xu,$^{21}$ L.~Yuan,$^{6}$ X.~Zheng,$^{5}$ H.Zhu$^{16}$}

\address{
$^{1}$ Yerevan Physics Institute, Yerevan, Armenia \\
$^{2}$ Rensselaer Polytechnic Institute, Troy, New York 12180 \\
$^{3}$ North Carolina A \& T State University, Greensboro, North Carolina 27411 \\
$^{4}$ Bucharest University, Bucharest, Romania \\
$^{5}$ Argonne National Laboratory, Argonne, Illinois 60439 \\
$^{6}$ Hampton University, Hampton, Virginia 23668 \\
$^{7}$ Thomas Jefferson National Accelerator Facility, Newport News, Virginia 23606 \\
$^{8}$ University of Maryland, College Park, Maryland 20742 \\
$^{9}$ The George Washington University, Washington, D.C. 20052 \\
$^{10}$ Vassar College, Poughkeepsie, New York 12604 \\
$^{11}$ Vrije Universiteit, 1081 HV Amsterdam, The Netherlands \\
$^{12}$ Florida International University, University Park, Florida 33199 \\
$^{13}$ University of Massachusetts Amherst, Amherst, Massachusetts 01003 \\
$^{14}$ University of the Witwatersrand, Johannesburg, South Africa \\
$^{15}$ University of Houston, Houston, TX 77204 \\
$^{16}$ University of Virginia, Charlottesville, Virginia 22901 \\
$^{17}$ Mississippi State University, Mississippi State, Mississippi 39762 \\
$^{18}$ Triangle Universities Nuclear Laboratory and Duke University, Durham, North Carolina 27708 \\
$^{19}$ California Institute of Technology, Pasadena, California 91125 \\
$^{20}$ University of North Carolina Wilmington, Wilmington, North Carolina 28403 \\ 
$^{21}$ University of Regina, Regina, Saskatchewan, Canada, S4S 0A2 \\
$^{22}$ Rutgers, The State University of New Jersey, Piscataway, New Jersey, 08855 \\
$^{23}$ University of Connecticut, Storrs, Connecticut 06269 \\
$^{24}$ Ohio University, Athens, Ohio 45071 \\
$^{25}$ James Madison University, Harrisonburg, Virginia 22807 \\
$^{26}$ Gettysburg College, Gettysburg, Pennsylvania 18103}
\newpage
\date{\today}

\begin{abstract}
A large data set of charged-pion ($\pi^\pm$) electroproduction from both
hydrogen and deuterium targets has been obtained spanning the low-energy
residual-mass region. These data conclusively show the onset of the
quark-hadron duality phenomenon, as predicted for high-energy hadron
electroproduction. We construct several ratios from these data to exhibit
the relation of this phenomenon to the high-energy factorization ansatz of
electron-quark scattering and subsequent quark $\rightarrow$ pion production
mechanisms.
\end{abstract}


\maketitle


At high energies, the property of Quantum Chromodynamics (QCD)
known as asymptotic freedom allows for an efficient
description in terms of quarks and gluons --- or partons, weakly interacting
at short distances. In contrast, at low energies the effects of confinement
impose a more efficient description in terms of collective degrees of freedom,
the physical mesons and baryons --- or hadrons.

Despite this apparent dichotomy, in nature there exist instances where
low-energy hadronic phenomena, averaged over appropriate energy intervals
\cite{PQW}, closely resemble those at asymptotically high energies, calculated
in terms of quark-gluon degrees of freedom. This is referred to as
quark-hadron duality, and reflects the relationship between the strong and
weak interaction limits of QCD --- confinement and asymptotic freedom.

The observation of this phenomenon in fact preceded QCD by a decade or so,
with remarkable similarity found between the
low-energy cross sections and high-energy behavior in hadronic reactions,
with the former on average appearing to mimic features of the latter. At
that time, this was explained with the development of Finite Energy Sum Rules,
relating dispersion integrals over resonance amplitudes at low energies to
Regge parameters describing the high-energy scattering \cite{FI77}.
The equivalence, on average, of hadron production in electron-positron
annihilation and the underlying quark-antiquark production mechanism was later
similarly understood \cite{BIGI_REV}.

It was natural, therefore, that this same framework was used to interpret
the early observation of quark-hadron duality
in inclusive electron-nucleon scattering.
Bloom \& Gilman found that by averaging the proton $F_2$ structure function
data over an appropriate energy range, the resulting structure function in
the resonance region closely resembled the scaling function which described
the high-energy scattering of electrons from point-like partons \cite{BG71}.
Recently, the phenomenon has been revisited with unprecedented precision,
and was found to work quantitatively far better, and far more locally,
than could have been expected \cite{NIC00,MEK05}.

Although postulated to be a general property of QCD, the dynamical origin of
quark-hadron duality remains poorly understood. It should manifest itself in
a wide variety of processes and observables. In this Letter, we generalize
the duality concept to the unexplored region of (``semi-inclusive'') pion
electroproduction \cite{CHE74,ACW00}, $eN \rightarrow e\pi^{\pm}X$, in which
a charged pion is detected in coincidence with a scattered electron.
The missing mass of the residual system $X$, $M_x$, is in the nucleon
resonance region (defined here as $M_x^2 <$ 4 GeV$^2$)
for the remainder of this Letter, and we will show
the dual behavior of this region with a high-energy parton description.

At high energies, perturbative QCD predicts factorization between the virtual
photon--quark interaction and the subsequent quark hadronization,
\begin{eqnarray}
{{{d\sigma} \over {d\Omega_e dE_{e^\prime} dz dp_T^2 d\phi}} \over
{{d\sigma} \over {d\Omega_e dE_{e^\prime}}}} = {{dN} \over {dz}} b e^{-bp_T^2}
{{1 + A cos(\phi) + B cos(2\phi)} \over {2\pi}}, \\
\label{eq:semi-parton}
{{dN} \over {dz}} \sim \sum_q e_q^2\ q(x,Q^2)\ D_{q \to \pi}(z,Q^2) ,
\end{eqnarray}
where the fragmentation function $D_{q \to \pi}(z,Q^2)$ gives the
probability for a quark to evolve into a pion $\pi$ detected
with a fraction $z$ of the quark (or virtual photon) energy, $z=E_{\pi}/\nu$.
The parton distribution functions $q(x,Q^2)$ are the usual functions depending
on the Bjorken variable $x$ and $Q^2$. The transverse momentum $p_T$, $z$ and
the angle $\phi$ reflect the extra kinematical degree of freedom associated
with the pion momentum. Both the parton distribution functions and the
fragmentation functions depend on $Q^2$ through logarithmic $Q^2$ evolution.
Their dependence on $p_T$ is removed in a Gaussian approximation, reflected
in the noted exponential $p_T$ dependence, with $b$ the average transverse
momentum of the
struck quark. In the (very) high energy limit, the factors $A$ and $B$ become
zero. At lower energies, these ``factors'' reflect the longitudinal-transverse
and transverse-transverse interference structure functions of the general
pion electroproduction framework \cite{RD88}, and can, {\sl e.g.}, vary with
$z$ and $Q^2$. Note that a
consequence of this factorization ansatz is that the fragmentation function is
independent of $x$, and the parton distribution function is independent of $z$.

At lower energies, where hadronic phenomena dominate, it is certainly
not obvious that the pion electroproduction
process factorizes in the same manner as in Eq.~(\ref{eq:semi-parton}).
However, it has been argued that at relatively low, yet sufficiently high
energies, with the quark-hadron duality phenomenon to occur, factorization
may still be possible \cite{IJMV01,CI01,MEK05}.

The experiment (E00-108) ran in the summer of 2003 in Hall C at Jefferson Lab.
An electron beam with a current ranging between 20 and 60 $\mu A$ was 
provided by the CEBAF accelerator with a beam energy of 5.5 GeV.
Incident electrons were scattered from a 4-cm-long liquid hydrogen or
deuterium target and detected in the Short Orbit Spectrometer (SOS).
The SOS central momentum remained constant throughout the experiment, with
a value of 1.7 GeV. The electroproduced mesons (predominantly pions) were
detected in the High Momentum Spectrometer (HMS), with momenta ranging from
1.3 to 4.1 GeV. The experiment consisted of two parts: i) at a fixed
electron kinematics of ($x,Q^2$) = (0.32, 2.30 GeV$^2$) the central HMS
momentum was varied to cover a range of 0.3 $< z <$ 1.0;
and ii) similarly, at $z$ = 0.55, the electron scattering angle was varied,
at constant momentum transfer angle, to span a range in $x$ from 0.22 to 0.58.
Note that this corresponds to an increase in $Q^2$, from 1.5 to 4.2 GeV$^2$.
The invariant mass squared, $W^2$, is typically 5.7 GeV$^2$ and always larger
than 4.2 GeV$^2$, well in the deep inelastic region, and all measurements were
performed for both $\pi^+$ and $\pi^-$.

Events from the aluminum walls of the cryogenic target cell were subtracted by
performing substitute empty target runs. Scattered electrons were selected by
the use of both a gas Cherenkov counter and an electromagnetic calorimeter.
Pions were selected using the coincidence time difference between scattered
electrons and secondary hadrons. In addition, an aerogel detector was used for
pion identification \cite{Asa05}. For kinematics with pion momenta above
2.4 GeV a correction was made to remove kaons from the pion sample, 10\% in
the worst case (at $z \sim$ 1), as determined from the electron-hadron
coincidence time.
From a measurement detecting positrons in SOS in coincidence with pions in HMS,
we found the background originating from $\pi^0$ production and its subsequent
decay into two photons and then electron-positron pairs, negligible.

We modelled semi-inclusive pion electroproduction \cite{gaskell}, following the
high-energy expectation of Eq.~(\ref{eq:semi-parton}). We used the CTEQ5
next-to-leading-order (NLO) parton distribution functions to parameterize
$q(x,Q^2)$ \cite{CTEQ}, and the fragmentation function parameterization for
$D^+_{q \rightarrow \pi}(z,Q^2) + D^-_{q \rightarrow \pi}(z,Q^2)$,
with $D^+$ ($D^-$) the favored (unfavored) fragmentation function, from
Binnewies {\sl et al.} \cite{BKK95}. The remaining unknowns are the ratio of
$D^-/D^+$, taken from a HERMES analysis \cite{Geiger}, the slope $b$ of the
$p_T$ dependence, and the factors $A$ and $B$ describing the $\phi$
dependence.

We can not constrain $b$ well within our own data set due to the limited
($p_T$,$\phi$) acceptance of a magnetic spectrometer setup. Here, with the
possible strong correlation between the $p_T$ and $\phi$ dependence
\cite{CAHN}, additional assumptions are required. Hence, we will use the
slope $b$ from an empirical fit to the HERMES $p_T$ dependence
($b \approx$ 4.66 GeV$^{-2}$) \cite{Hommez}. Our own best estimate is $b$ = 4.0
$\pm$ 0.4 GeV$^{-2}$, with no noticable differences between $b$-values
extracted from the $p_T$-dependence of either $\pi^+$ and $\pi^-$ data,
or $^1$H and $^2$H data, somewhat lower than the HERMES slope.
We do find a $\phi$ dependence in our data, with typical parameters of
$A$ = 0.16 $\pm$ 0.04, and $B$ = 0.02 $\pm$ 0.02, for an average $<p_T>$ =
0.1 GeV. These $\phi$-dependences become smaller to negligible in the ratios of
cross sections shown later. Similarly, we find a $Q^2$-dependence in our
data that differs from the factorized high-energy
expectation, but this does not affect the results shown
below. Of course, these findings do cast doubt on the strict applicability
of the high-energy approximation for our experiment.

Within our Monte Carlo package, we estimated two non-trivial corrections
to the data. Radiative corrections were applied in two steps. We directly
estimated the radiation tails within our semi-inclusive pion electroproduction
data using the Monte Carlo. In addition, we explicitly
subtracted radiation tails coming from the exclusive reactions
$e + p \rightarrow e^\prime + \pi^+ + n$ and
$e + n \rightarrow e^\prime + \pi^- + p$. For these processes, we interpolated
between the low-$W^2$, low-$Q^2$ predictions using the MAID model \cite{MAID}
and the higher-$W^2$ data of Brauel {\sl et al.} and Bebek {\sl et al.}
\cite{Bra79,Beb78}.
We subtracted events from diffractive $\rho$ production, using PYTHIA
\cite{pythia} to estimate the p(e,e$^\prime\rho^\circ$)p cross section
with similar
modifications as implemented by the HERMES collaboration \cite{Hommez,Tytgat}.
We also made a 2\% correction to the deuterium data to account
for the loss of pions traversing the deuterium nucleus \cite{Sar05}.

\begin{figure}
\begin{center}
\scalebox{1.0}[1.0]{\includegraphics{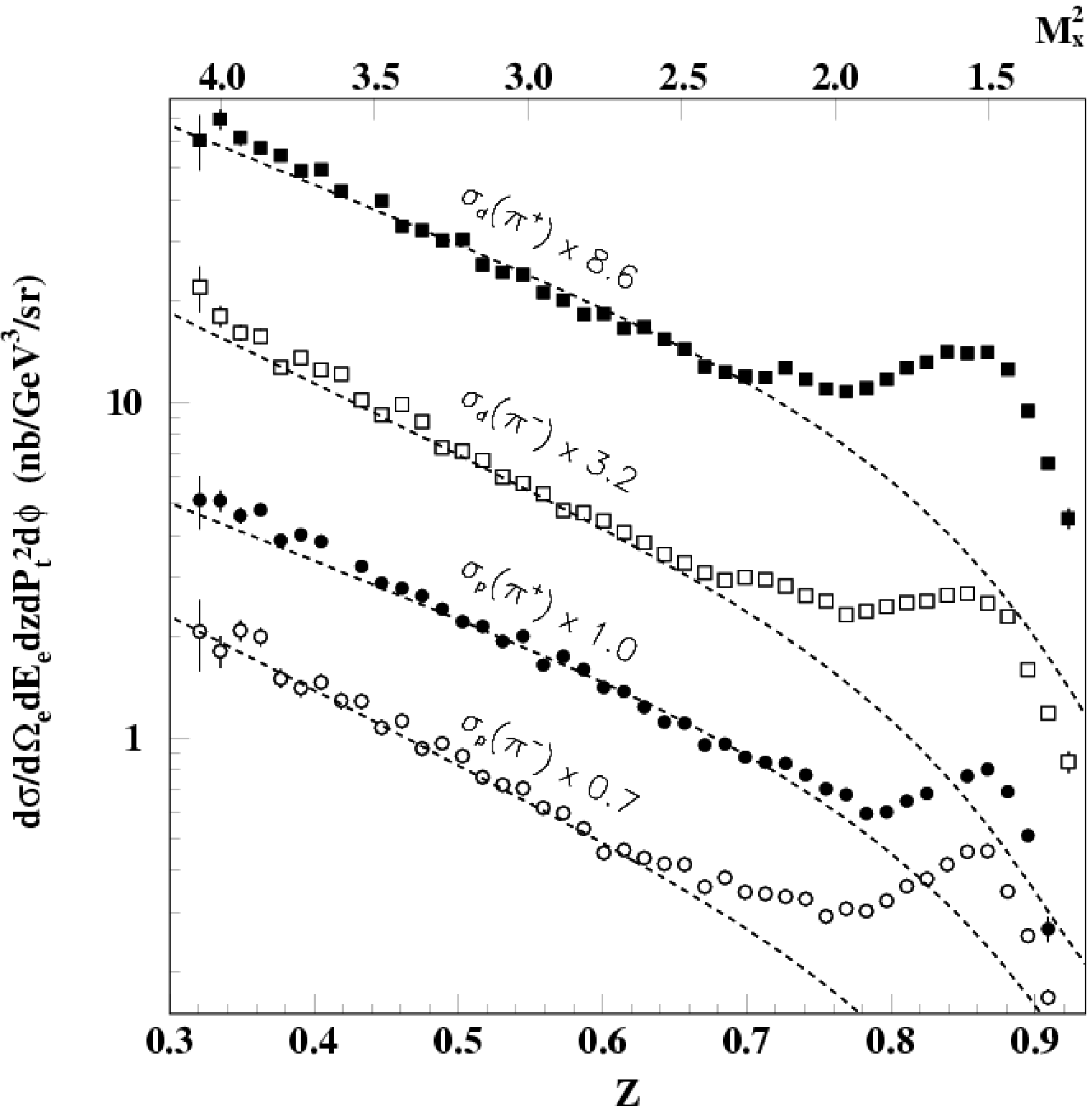}}
\caption{\label{fig:hpiminus}
The $^{1,2}$H(e,e$^\prime \pi^\pm$)X
cross sections at $x$ = 0.32 as a function of $z$ 
in comparison with Monte Carlo simulations
(dashed curves) starting from a fragmentation ansatz (see text).
The various cross sections have been multiplied as indicated for
the purpose of plotting.
See Table~I for numerical values.}
\end{center}
\end{figure}

The $^{1,2}$H(e,e$^\prime \pi^\pm$)$X$ cross sections as measured at
$x$ = 0.32 are compared with the results of the simulation in
Fig.~\ref{fig:hpiminus}, as a function of $z$. The general agreement between
data and Monte Carlo is excellent for $z <$ 0.65. Within our kinematics
($p_T \sim$ 0), $M_x^2$ is almost directly related to $z$, as $M_x^2 \approx
M_p^2 + Q^2(1/x-1)(1-z)$. Hence, the large excess at $z >$ 0.8 in the data with
respect to the simulation mainly reflects the $\gamma N-\pi\Delta(1232)$
transition region. Indeed, in {\sl e.g.} a typical $^1$H(e,e$^\prime \pi^-$)$X$
spectrum one can see one prominent $\Delta(1232)$ resonance, and only some
small structure beyond \cite{Bra79,Beb78}.
Apparently, above $M_x^2$ = 2.5 GeV$^2$ or so, there are already sufficient
resonances to render a spectrum mimicking the smooth $z$-dependence as
expected from the Monte Carlo simulation following the factorization ansatz
of Eq.~(\ref{eq:semi-parton}). Lastly, the fast drop of the simulations at
large $z$ may be artificial. Whereas fragmentation functions have
been well mapped up to $z$ = 0.9 at the LEP collider \cite{AKK05}, to better
than 50\%, there remain questions for semi-inclusive pion production at lower
$Q^2$. Here, the fragmentation functions could well flatten out \cite{Dre78},
as also included in the Field and Feynman expectations \cite{FF78}, that tend
to produce more particles at lower energies beyond $z$ = 0.7 or so.

To quantify the surprising resemblance of semi-inclusive pion electroproduction
data in the nucleon resonance region with the high energy prediction of
Eq.~(\ref{eq:semi-parton}), we formed simple ratios of the measured
cross sections, insensitive to the fragmentation process (assuming charge
symmetry) at leading order (LO) in $\alpha_s$. If one neglects strange quarks
and any $p_T$-dependence to the parton distribution functions, these ratios can
be expressed in terms of $u$ and $d$ parton distributions, as follows
\begin{eqnarray}
\label{eq:fact}
{{\sigma_p(\pi^+) + \sigma_p(\pi^-)} \over {\sigma_d(\pi^+) + \sigma_d(\pi^-)}}
  = {{4u(x) + 4{\bar u}(x) + d(x) + {\bar d}(x)} \over 
    {5(u(x) + d(x) + {\bar u}(x) + {\bar d}(x))}}, \\
{{\sigma_p(\pi^+) - \sigma_p(\pi^-)} \over {\sigma_d(\pi^+) - \sigma_d(\pi^-)}}
  = {{4u_v(x) - d_v(x)} \over {3(u_v(x) + d_v(x))}},
\end{eqnarray}
with the notation $\sigma_p(\pi^+)$ refering to the $\pi^+$ pion
electroproduction cross section off the proton, $u = u_v + {\bar u}$,
$d = d_v + {\bar d}$, and the $Q^2$-dependence left
out of these formulas for convenience. These ratios allow us to study the
factorization ansatz in more detail, with both ratios rendering results
independent of $z$ (and $p_T$).

\begin{figure}
\begin{center}
\scalebox{1.0}[1.0]{\includegraphics{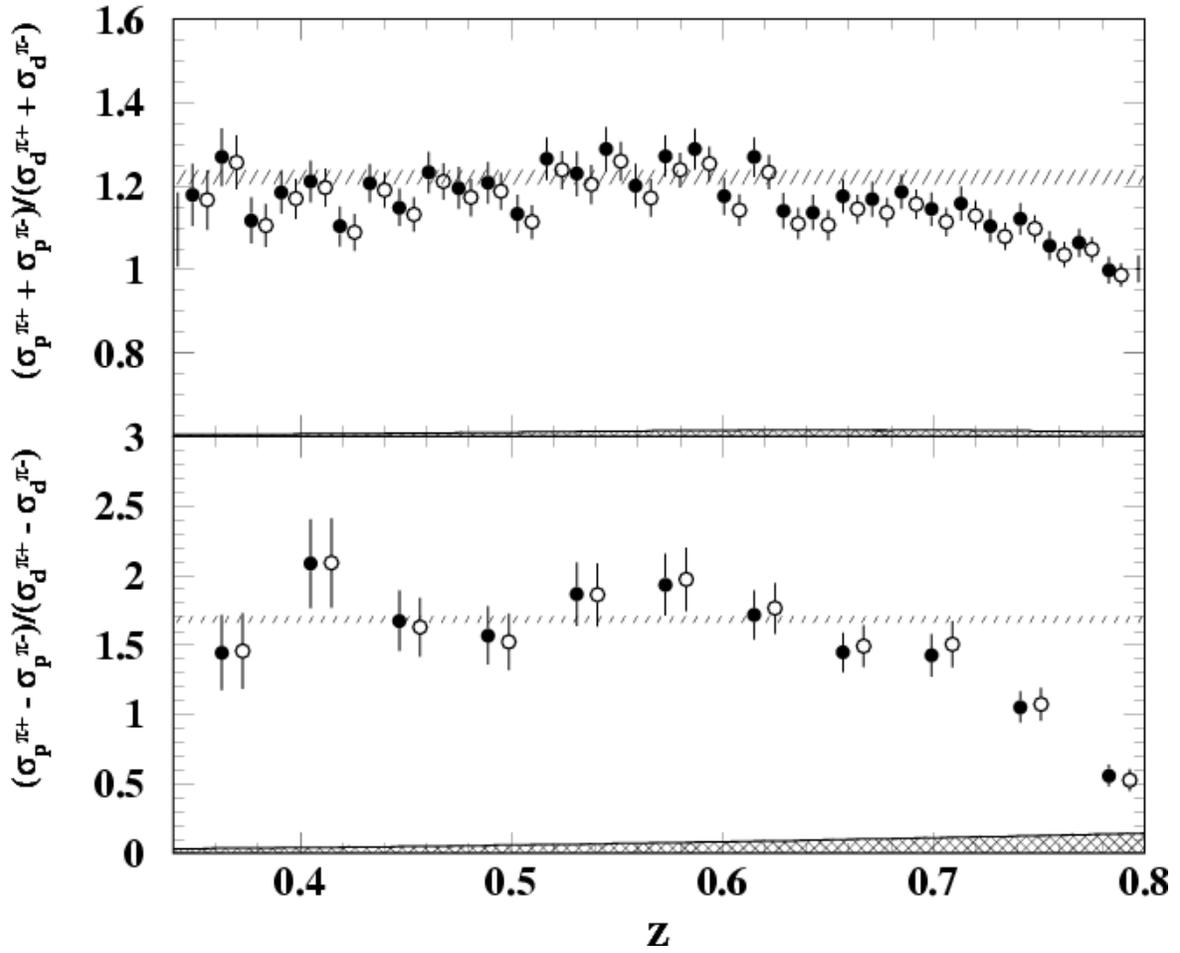}}
\caption{\label{fig:factorization}
The ratio of proton to deuterium results of the sum (top) and
difference (bottom) of $\pi^+$ and $\pi^-$ cross sections as a function
of $z$, at $x$ = 0.32.
Closed (open) symbols reflect data after (before)
events from coherent $\rho$ production are subtracted (see text). 
The symbols have been sightly offset in $z$ for clarity.
The hatched areas in the bottom indicate the systematic 
uncertainties, whereas
the shaded bands represent a variety of calculations,
 at both leading order
and next-to-leading-order of $\alpha_s$, of the shown ratio
\protect\cite{grv98,CTEQ}.
See Tables~II and III for numerical values.}
\end{center}
\end{figure}

We show our results in Fig.~\ref{fig:factorization}, with
the solid (open) symbols reflecting the data after (before) subtraction of
the diffractive $\rho$ contributions. The hatched areas in the bottom
indicate the estimated systematic uncertainty. The shaded bands reflect
the expectations under the assumptions described above (factorization, no
strange quark effects, charge symmetry for the fragmentation functions),
and include a variety of calculations, using both LO and NLO ($M\bar{S}$
and valence) parton distribution functions from the GRV collaboration, and
NLO calculations from the CTEQ collaboration \cite{grv98,CTEQ}.

Our data are remarkably close to the near-independence of $z$ as expected
in the high-energy limit, with the clearest deviations in the region of
$z >$ 0.7, approaching on the $\Delta(1232)$ residual mass region.
Within 10\% we find perfect agreement beyond this region.

\begin{figure}
\begin{center}
\scalebox{1.0}[1.0]{\includegraphics{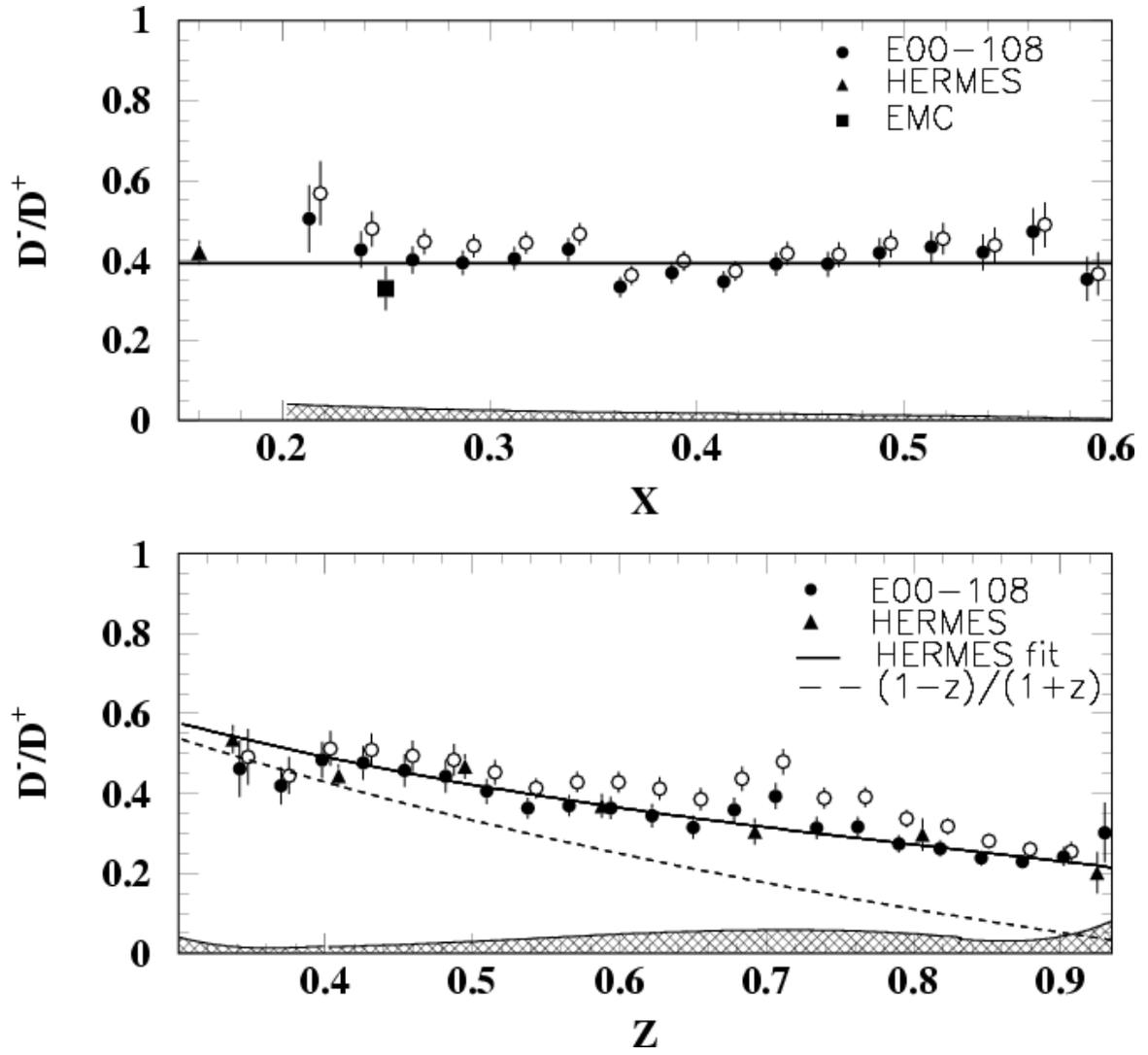}}
\vspace{0.5cm}
\caption{\label{fig:dminusdplus}
{\sl Top:} The ratio of unfavored to favored fragmentation 
function $D^-/D^+$ as a function of $x$ at $z$ = 0.55, 
evaluated at leading order of $\alpha_s$
from the deuterium data. The triangles (square) reflect HERMES (EMC) data
\protect\cite{Hommez,EMC}, with the solid curve a fit to HERMES data. 
Further symbols and the hatched area are as in Fig. 2.
{\sl Bottom:} Same as {\sl top}, but now as 
a function of $z$ for $x$ = 0.32.
The dashed curve represents the expectation \protect\cite{FF78}
under the independent fragmentation hypothesis.
See Tables~IV and V  for numerical values.}
\end{center}
\end{figure}

Using the deuterium data only, the ratio of unfavored to favored fragmentation
functions $D^-/D^+$ can be extracted. This ratio is, to a good approximation,
at LO simply given by 
\begin{eqnarray}
\label{eq:dmindplus}
{D^-}/{D^+} = 
\left ( {4 - {{\sigma_d(\pi^+)} \over {\sigma_d(\pi^-)}}} \right ) {\big /}
\left ( {4~{{\sigma_d(\pi^+)} \over {\sigma_d(\pi^-)}} - 1} \right ).
\end{eqnarray}
In the high-energy limit, this ratio should solely depend on $z$ (and $Q^2$),
but not on $x$. The results are shown in Fig.~\ref{fig:dminusdplus},
with the closed (open) symbols reflecting the data after (before) subtraction
of the diffractive $\rho$ contributions. The solid curves are a fit to
the HERMES data for the same ratio \cite{Geiger}. The dashed curve is the
expectation $(1-z)/(1+z)$ according to Field and Feynman for independent
fragmentation \cite{FF78}. The hatched areas indicate the systematic
uncertainties, dominated by uncertainties due to the two non-trivial
corrections discussed above.

We observe that the extracted values for $D^-/D^+$ closely resemble those
of the HERMES experiment \cite{Geiger}. The data show a near-independence
as a function of $x$, as expected from Eq.~(\ref{eq:semi-parton}),
and a smooth slope as a function of $z$, reflecting a fit to the higher-energy
HERMES data, all at $M_x^2 > 4$ GeV$^2$. This is quite remarkable given that
our data cover the full resonance region for the residual system $X$,
$M_p^2 < M_x^2 < 4.2$~GeV$^2$. Apparently, there is a mechanism at work that
removes the resonance excitations in the $\pi^+/\pi^-$ ratio, and hence the
$D^-/D^+$ ratio. We note that both our data and the fit to the higher-energy
HERMES data far exceed the Field and Feynman expectations for large $z$.

The mechanism above can be simply understood in the SU(6) symmetric quark
model. Close \& Isgur \cite{CI01} applied this to calculate production rates
in various channels in semi-inclusive pion photoproduction,
$\gamma N \to \pi X$.
The pattern of constructive and destructive interference, which was
a crucial feature of the appearance of duality in inclusive structure
functions, is in this model also repeated in the semi-inclusive case.
The results suggest an explanation for the smooth behavior of
$D^-/D^+ \equiv D_d^{\pi^+}/D_u^{\pi^+}$ for a deuterium target in
Fig.~\ref{fig:dminusdplus}. The relative weights of the
photoproduction matrix elements, summed over $p$ and $n$, is for
$\pi^+$ production always 4 times larger than for $\pi^-$ production.
In the SU(6) limit, therefore, the resonance contributions to the
ratio of Eq.~(\ref{eq:dmindplus}) cancel exactly, leaving behind only the
smooth background, as would be expected at high energies.
This may account for the glaring lack of resonance structure in
the resonance region fragmentation functions in Fig.~\ref{fig:dminusdplus}.

In summary, we have measured charged-pion ($\pi^{\pm}$) electroproduction cross
sections for both hydrogen and deuterium targets. Our data cover the region
where the missing mass of the residual system $X$ is in the resonance region.
We observe for the first time the quark-hadron duality phenomenon in such
reactions, in that such data equate the high-energy expectations. We have
quantified this behavior by constructing several ratios from these data, that
exhibit, at low energies, the features of factorization in an electron-quark
scattering and a subsequent quark-pion fragmentation process. Furthermore,
the ratio of favored to unfavored fragmentation functions closely resembles
that of high energy reactions, over the full range of missing mass.
This observation can be explained in the $SU(6)$ symmetric quark model.

\medskip
The authors wish to thank A. Bruell, C.E. Carlson and W. Melnitchouk for
helpful discussions. This work is supported in part by research grants from the
U.S. Department of Energy, the U.S. National Science Foundation, the Natural
Sciences and Engineering Research Council of Canada, and FOM (Netherlands).
The Southeastern Universities Research Association operates the
Thomas Jefferson National Accelerator Facility under the
U.S. Department of Energy contract DEAC05-84ER40150.

\begin{table}[htbp]
  \begin{center}
    \begin{tabular}{ccccc}
    \hline \hline \small \small
$z$ & $\sigma_p(\pi^+)$ & $\sigma_p(\pi^-)$ &
    $\sigma_d(\pi^+)$ & $\sigma_d(\pi^-)$ \\
\hline
 0.321 & $5.1022\pm 0.9260$ & $2.9520\pm 0.6990$ & $6.9866\pm 1.2920$ & $6.8482\pm 3.3984$ \\
 0.335 & $5.0775\pm 0.3760$ & $2.5860\pm 0.2920$ & $8.0795\pm 0.5360$ & $5.6131\pm 1.2928$ \\
 0.349 & $4.5875\pm 0.2620$ & $2.9750\pm 0.2130$ & $7.1233\pm 0.3650$ & $5.0146\pm 0.8896$ \\
 0.363 & $4.7595\pm 0.2211$ & $2.8503\pm 0.1767$ & $6.6510\pm 0.3062$ & $4.8947\pm 0.7571$ \\
 0.377 & $3.8615\pm 0.1674$ & $2.1376\pm 0.1349$ & $6.3132\pm 0.2364$ & $3.9476\pm 0.5645$ \\
 0.391 & $4.0219\pm 0.1484$ & $2.0030\pm 0.1192$ & $5.6805\pm 0.2063$ & $4.2287\pm 0.5348$ \\
 0.405 & $3.8327\pm 0.1381$ & $2.0858\pm 0.1150$ & $5.7105\pm 0.1963$ & $3.8901\pm 0.4964$ \\
 0.419 & $3.2211\pm 0.1261$ & $1.8373\pm 0.1061$ & $4.9272\pm 0.1762$ & $3.7896\pm 0.4676$ \\
 0.433 & $3.2277\pm 0.1103$ & $1.8290\pm 0.0876$ & $4.8902\pm 0.1742$ & $3.1669\pm 0.3816$ \\
 0.447 & $2.8839\pm 0.1027$ & $1.5252\pm 0.0765$ & $4.6133\pm 0.1460$ & $2.8639\pm 0.3530$ \\
 0.461 & $2.7786\pm 0.1018$ & $1.6006\pm 0.0734$ & $3.8764\pm 0.1357$ & $3.0718\pm 0.3539$ \\
 0.475 & $2.6399\pm 0.1016$ & $1.3224\pm 0.0671$ & $3.7653\pm 0.1338$ & $2.7262\pm 0.3348$ \\
 0.489 & $2.4154\pm 0.0954$ & $1.3694\pm 0.0610$ & $3.5153\pm 0.1226$ & $2.2797\pm 0.2849$ \\
 0.503 & $2.2102\pm 0.0884$ & $1.2570\pm 0.0533$ & $3.5419\pm 0.1153$ & $2.2285\pm 0.2474$ \\
 0.517 & $2.1440\pm 0.0832$ & $1.0879\pm 0.0461$ & $2.9755\pm 0.1064$ & $2.0947\pm 0.2295$ \\
 0.531 & $1.9303\pm 0.0812$ & $1.0323\pm 0.0441$ & $2.8307\pm 0.1048$ & $1.8624\pm 0.2091$ \\
 0.545 & $2.0077\pm 0.0799$ & $1.0071\pm 0.0410$ & $2.7786\pm 0.1125$ & $1.7867\pm 0.1958$ \\
 0.559 & $1.6436\pm 0.0720$ & $0.8803\pm 0.0367$ & $2.4638\pm 0.0944$ & $1.6663\pm 0.1816$ \\
 0.573 & $1.7433\pm 0.0643$ & $0.8498\pm 0.0319$ & $2.3350\pm 0.0800$ & $1.4832\pm 0.1527$ \\
 0.587 & $1.5925\pm 0.0566$ & $0.7649\pm 0.0282$ & $2.1198\pm 0.0734$ & $1.4626\pm 0.1479$ \\
 0.601 & $1.4067\pm 0.0504$ & $0.6474\pm 0.0251$ & $2.1260\pm 0.0714$ & $1.3834\pm 0.1430$ \\
 0.615 & $1.3711\pm 0.0478$ & $0.6604\pm 0.0244$ & $1.9290\pm 0.0660$ & $1.2753\pm 0.1339$ \\
 0.629 & $1.2301\pm 0.0441$ & $0.6232\pm 0.0236$ & $1.9385\pm 0.0636$ & $1.1877\pm 0.1289$ \\
 0.643 & $1.1113\pm 0.0393$ & $0.5993\pm 0.0217$ & $1.7834\pm 0.0573$ & $1.0988\pm 0.1162$ \\
 0.657 & $1.1037\pm 0.0365$ & $0.5978\pm 0.0197$ & $1.6701\pm 0.0508$ & $1.0364\pm 0.1036$ \\
 0.671 & $0.9497\pm 0.0322$ & $0.5131\pm 0.0180$ & $1.4767\pm 0.0459$ & $0.9665\pm 0.0979$ \\
 0.685 & $0.9577\pm 0.0320$ & $0.5469\pm 0.0191$ & $1.4249\pm 0.0432$ & $0.9206\pm 0.0923$ \\
 0.699 & $0.8742\pm 0.0296$ & $0.4948\pm 0.0177$ & $1.3854\pm 0.0427$ & $0.9392\pm 0.0925$ \\
 0.713 & $0.8428\pm 0.0286$ & $0.4895\pm 0.0177$ & $1.3738\pm 0.0423$ & $0.9225\pm 0.0924$ \\
 0.727 & $0.8349\pm 0.0283$ & $0.4788\pm 0.0171$ & $1.4706\pm 0.0431$ & $0.8840\pm 0.0887$ \\
 0.741 & $0.7730\pm 0.0261$ & $0.4720\pm 0.0165$ & $1.3586\pm 0.0396$ & $0.8252\pm 0.0821$ \\
 0.755 & $0.7029\pm 0.0222$ & $0.4195\pm 0.0147$ & $1.2689\pm 0.0345$ & $0.7955\pm 0.0732$ \\
 0.769 & $0.6755\pm 0.0216$ & $0.4409\pm 0.0156$ & $1.2450\pm 0.0336$ & $0.7247\pm 0.0674$ \\
 0.783 & $0.5922\pm 0.0192$ & $0.4336\pm 0.0145$ & $1.2769\pm 0.0340$ & $0.7418\pm 0.0674$ \\
 0.797 & $0.5987\pm 0.0194$ & $0.4654\pm 0.0155$ & $1.3623\pm 0.0360$ & $0.7660\pm 0.0687$ \\
 0.811 & $0.6461\pm 0.0192$ & $0.5136\pm 0.0163$ & $1.4704\pm 0.0375$ & $0.7862\pm 0.0692$ \\
 0.825 & $0.6808\pm 0.0199$ & $0.5425\pm 0.0166$ & $1.5252\pm 0.0375$ & $0.7965\pm 0.0697$ \\
 0.839 & $0.7641\pm 0.0213$ & $0.5979\pm 0.0177$ & $1.6375\pm 0.0397$ & $0.8278\pm 0.0698$ \\
 0.853 & $0.7667\pm 0.0212$ & $0.6530\pm 0.0186$ & $1.6277\pm 0.0369$ & $0.8383\pm 0.0670$ \\
 0.867 & $0.8048\pm 0.0210$ & $0.6553\pm 0.0171$ & $1.6313\pm 0.0341$ & $0.7840\pm 0.0598$ \\
 0.881 & $0.6894\pm 0.0212$ & $0.4966\pm 0.0144$ & $1.4526\pm 0.0311$ & $0.7161\pm 0.0574$ \\
 0.895 & $0.5098\pm 0.0209$ & $0.3664\pm 0.0131$ & $1.0932\pm 0.0279$ & $0.4969\pm 0.0500$ \\
 0.909 & $0.2692\pm 0.0232$ & $0.2400\pm 0.0122$ & $0.7644\pm 0.0274$ & $0.3685\pm 0.0512$ \\
 0.923 & $0.0222\pm 0.0350$ & $0.1380\pm 0.0120$ & $0.5219\pm 0.0365$ & $0.2641\pm 0.0630$ \\
    \hline \hline
    \end{tabular}
  \end{center}
  \caption{Cross sections as a function of $z$ in nb/GeV$^3$/sr
coresponding to Fig.~1. Errors are statistical only.}
\end{table}

\begin{table}[htbp]
  \begin{center}
    \begin{tabular}{ccccc}
    \hline \hline
$z$ & $R_{pd}^+$ (after $\rho$) & $R_{pd}^+$ (before $\rho$)  \\ 
\hline
 0.349 & $1.1790\pm 0.0750$ & $1.1670\pm 0.0710$ \\
 0.363 & $1.2700\pm 0.0700$ & $1.2570\pm 0.0650$ \\
 0.377 & $1.1180\pm 0.0550$ & $1.1060\pm 0.0510$ \\
 0.391 & $1.1850\pm 0.0510$ & $1.1700\pm 0.0480$ \\
 0.405 & $1.2110\pm 0.0500$ & $1.1970\pm 0.0460$ \\
 0.419 & $1.1040\pm 0.0480$ & $1.0900\pm 0.0440$ \\
 0.433 & $1.2070\pm 0.0460$ & $1.1910\pm 0.0420$ \\
 0.447 & $1.1490\pm 0.0450$ & $1.1320\pm 0.0410$ \\
 0.461 & $1.2330\pm 0.0490$ & $1.2120\pm 0.0440$ \\
 0.475 & $1.1960\pm 0.0510$ & $1.1730\pm 0.0450$ \\
 0.489 & $1.2090\pm 0.0500$ & $1.1880\pm 0.0440$ \\
 0.503 & $1.1340\pm 0.0460$ & $1.1140\pm 0.0400$ \\
 0.517 & $1.2660\pm 0.0520$ & $1.2390\pm 0.0460$ \\
 0.531 & $1.2300\pm 0.0530$ & $1.2040\pm 0.0470$ \\
 0.545 & $1.2890\pm 0.0540$ & $1.2600\pm 0.0470$ \\
 0.559 & $1.2010\pm 0.0530$ & $1.1720\pm 0.0460$ \\
 0.573 & $1.2720\pm 0.0490$ & $1.2390\pm 0.0420$ \\
 0.587 & $1.2890\pm 0.0490$ & $1.2540\pm 0.0420$ \\
 0.601 & $1.1760\pm 0.0450$ & $1.1430\pm 0.0380$ \\
 0.615 & $1.2700\pm 0.0480$ & $1.2340\pm 0.0410$ \\
 0.629 & $1.1410\pm 0.0430$ & $1.1110\pm 0.0370$ \\
 0.643 & $1.1370\pm 0.0420$ & $1.1070\pm 0.0360$ \\
 0.657 & $1.1760\pm 0.0400$ & $1.1450\pm 0.0340$ \\
 0.671 & $1.1690\pm 0.0420$ & $1.1370\pm 0.0350$ \\
 0.685 & $1.1870\pm 0.0410$ & $1.1570\pm 0.0350$ \\
 0.699 & $1.1450\pm 0.0400$ & $1.1140\pm 0.0340$ \\
 0.713 & $1.1590\pm 0.0410$ & $1.1300\pm 0.0350$ \\
 0.727 & $1.1050\pm 0.0390$ & $1.0800\pm 0.0330$ \\
 0.741 & $1.1220\pm 0.0380$ & $1.0980\pm 0.0330$ \\
 0.755 & $1.0580\pm 0.0340$ & $1.0360\pm 0.0300$ \\
 0.769 & $1.0650\pm 0.0340$ & $1.0480\pm 0.0300$ \\
 0.783 & $0.9990\pm 0.0320$ & $0.9870\pm 0.0280$ \\
    \hline \hline
    \end{tabular}
  \end{center}
  \caption{The ratio $R_{pd}^+$ of proton to deuterium 
results of the sum 
of $\pi^+$ and $\pi^-$ cross sections as a function
of $z$, at $x$ = 0.32. Results 
reflect data after (left) or before (right) 
events from coherent $\rho$ production are subtracted 
(see text). Errors are statistical only. 
The systematic error is given by 
$0.0854 - 0.546 z + 1.18 z^2 - 0.778 z^3$ for $0.3<z<0.68$ and by
$-3.313 + 13.43 z - 17.99 z^2 + 7.995 z^3$ for $0.68<z<0.80$.
}
\end{table}

\begin{table}[htbp]
  \begin{center}
    \begin{tabular}{ccccc}
    \hline \hline
$z$ & $R_{pd}^-$ (after $\rho$) & $R_{pd}^-$ (before $\rho$)  \\ 
\hline
 0.363 & $1.4440\pm 0.2710$ & $1.4570\pm 0.2730$ \\
 0.405 & $2.0860\pm 0.3190$ & $2.0910\pm 0.3210$ \\
 0.447 & $1.6750\pm 0.2180$ & $1.6300\pm 0.2120$ \\
 0.489 & $1.5690\pm 0.2080$ & $1.5210\pm 0.2030$ \\
 0.531 & $1.8670\pm 0.2290$ & $1.8610\pm 0.2280$ \\
 0.573 & $1.9340\pm 0.2220$ & $1.9720\pm 0.2280$ \\
 0.615 & $1.7160\pm 0.1770$ & $1.7640\pm 0.1830$ \\
 0.657 & $1.4470\pm 0.1440$ & $1.4930\pm 0.1500$ \\
 0.699 & $1.4260\pm 0.1540$ & $1.5070\pm 0.1680$ \\
 0.741 & $1.0530\pm 0.1120$ & $1.0740\pm 0.1180$ \\
 0.783 & $0.5590\pm 0.0780$ & $0.5270\pm 0.0800$ \\
    \hline \hline
    \end{tabular}
  \end{center}
  \caption{The ratio $R_{pd}^-$ of proton to deuterium 
results of the difference
of $\pi^+$ and $\pi^-$ cross sections as a function
of $z$, at $x$ = 0.32. Results 
reflect data after (left) or before (right) 
events from coherent $\rho$ production are 
subtracted (see text). Errors are statistical only.
The systematic error is given by 
$0.046 - 0.203 z + 0.538 z^2 - 0.163 z^3$.}
\end{table}

\begin{table}[htbp]
  \begin{center}
    \begin{tabular}{ccccc}
    \hline \hline
$z$ & $D^-/D^+$ (after $\rho$) & $D^-/D^+$ (before $\rho$)  \\ 
\hline
 0.342 & $0.4620\pm 0.0710$ & $0.4920\pm 0.0700$ \\
 0.370 & $0.4196\pm 0.0475$ & $0.4449\pm 0.0465$ \\
 0.398 & $0.4838\pm 0.0453$ & $0.5126\pm 0.0438$ \\
 0.426 & $0.4764\pm 0.0429$ & $0.5087\pm 0.0411$ \\
 0.454 & $0.4575\pm 0.0414$ & $0.4940\pm 0.0392$ \\
 0.482 & $0.4425\pm 0.0413$ & $0.4837\pm 0.0395$ \\
 0.510 & $0.4059\pm 0.0318$ & $0.4530\pm 0.0306$ \\
 0.538 & $0.3635\pm 0.0270$ & $0.4134\pm 0.0257$ \\
 0.566 & $0.3699\pm 0.0266$ & $0.4288\pm 0.0253$ \\
 0.594 & $0.3638\pm 0.0274$ & $0.4280\pm 0.0267$ \\
 0.622 & $0.3448\pm 0.0298$ & $0.4124\pm 0.0284$ \\
 0.650 & $0.3157\pm 0.0289$ & $0.3853\pm 0.0279$ \\
 0.678 & $0.3587\pm 0.0314$ & $0.4376\pm 0.0307$ \\
 0.706 & $0.3934\pm 0.0327$ & $0.4800\pm 0.0319$ \\
 0.734 & $0.3137\pm 0.0273$ & $0.3889\pm 0.0264$ \\
 0.762 & $0.3164\pm 0.0254$ & $0.3911\pm 0.0254$ \\
 0.790 & $0.2738\pm 0.0223$ & $0.3375\pm 0.0223$ \\
 0.818 & $0.2625\pm 0.0198$ & $0.3177\pm 0.0198$ \\
 0.846 & $0.2380\pm 0.0177$ & $0.2808\pm 0.0168$ \\
 0.874 & $0.2294\pm 0.0161$ & $0.2607\pm 0.0159$ \\
 0.902 & $0.2423\pm 0.0243$ & $0.2555\pm 0.0238$ \\
 0.930 & $0.3025\pm 0.0739$ & $0.2944\pm 0.0724$ \\
    \hline \hline
    \end{tabular}
  \end{center}
  \caption{The ratio of unfavored to favored 
fragmentation function $D^-/D^+$
as a function of $z$ at $x$ = 0.32, evaluated at 
leading order of $\alpha_s$ from the deuterium data. Results 
reflect data after (left) or before (right) 
events from coherent $\rho$ production are subtracted (see text). 
Errors are statistical only.
The systematic error is given by 
$2.445 -18.434 z + 46.196 z^2 - 38.194 z^3$ for $0.2<z<0.3$,
$0.273 - 1.684 z + 3.461 z^2 - 2.132 z^3$ for $0.4<z<0.835$, and
$-24.15 + 89.532 z - 110.09 z^2 + 44.974 z^3$ for $0.835<z<0.935$.}
\end{table}

\begin{table}[htbp]
  \begin{center}
    \begin{tabular}{ccccc}
    \hline \hline
$x$ & $D^-/D^+$ (after $\rho$) & $D^-/D^+$ (before $\rho$)  \\ 
\hline
 0.213 & $0.5048\pm 0.0835$ & $0.5682\pm 0.0800$ \\
 0.238 & $0.4272\pm 0.0458$ & $0.4789\pm 0.0435$ \\
 0.263 & $0.4008\pm 0.0341$ & $0.4472\pm 0.0322$ \\
 0.287 & $0.3939\pm 0.0311$ & $0.4361\pm 0.0294$ \\
 0.312 & $0.4049\pm 0.0289$ & $0.4446\pm 0.0277$ \\
 0.338 & $0.4278\pm 0.0285$ & $0.4660\pm 0.0274$ \\
 0.363 & $0.3334\pm 0.0252$ & $0.3631\pm 0.0242$ \\
 0.388 & $0.3690\pm 0.0263$ & $0.3987\pm 0.0253$ \\
 0.413 & $0.3476\pm 0.0262$ & $0.3732\pm 0.0249$ \\
 0.438 & $0.3914\pm 0.0298$ & $0.4177\pm 0.0287$ \\
 0.463 & $0.3907\pm 0.0320$ & $0.4142\pm 0.0310$ \\
 0.488 & $0.4198\pm 0.0362$ & $0.4420\pm 0.0349$ \\
 0.513 & $0.4336\pm 0.0403$ & $0.4546\pm 0.0395$ \\
 0.538 & $0.4202\pm 0.0454$ & $0.4385\pm 0.0440$ \\
 0.562 & $0.4721\pm 0.0581$ & $0.4890\pm 0.0572$ \\
 0.588 & $0.3533\pm 0.0553$ & $0.3668\pm 0.0539$ \\
    \hline \hline
    \end{tabular}
  \end{center}
  \caption{The ratio of unfavored to favored 
fragmentation function $D^-/D^+$
as a function of $x$ at $z$ = 0.55, evaluated at 
leading order of $\alpha_s$ from the deuterium data. Results 
reflect data after (left) or before (right) 
events from coherent $\rho$ production are 
subtracted (see text). Errors are statistical only.
The systematic error is given by 
$0.126 - 0.669 x + 1.445 x^2 - 1.119 x^3.$
}
\end{table}

\end{document}